\documentstyle[12pt]{article}

\def\hybrid{\topmargin -20pt    \oddsidemargin 0pt  
        \headheight 0pt \headsep 0pt  
        \textwidth 6.25in       
        \textheight 9.5in       
        \marginparwidth .875in  
        \parskip 5pt plus 1pt   \jot = 1.5ex}  

\def\noi{\noindent}  
  
\def\baselinestretch{1.2}  
  
\catcode`\@=11  
  
\def\marginnote#1{}  
\def\draftlabel#1{{\@bsphack\if@filesw {\let\thepage\relax  
   \xdef\@gtempa{\write\@auxout{\string  
      \newlabel{#1}{{\@currentlabel}{\thepage}}}}}\@gtempa  
   \if@nobreak \ifvmode\nobreak\fi\fi\fi\@esphack}  
        \gdef\@eqnlabel{#1}}  
\def\@eqnlabel{}  
\def\@vacuum{}  
\def\draftmarginnote#1{\marginpar{\raggedright\scriptsize\tt#1}}  
  
\def\draft{\oddsidemargin -.2truein  
        \def\@oddfoot{\sl preliminary draft \hfil  
        \rm\thepage\hfil\sl\today\quad\militarytime}  
        \let\@evenfoot\@oddfoot \overfullrule 3pt  
        \let\label=\draftlabel  
        \let\marginnote=\draftmarginnote  
   \def\@eqnnum{(\theequation)\rlap{\kern\marginparsep\tt\@eqnlabel}%
\global\let\@eqnlabel\@vacuum}  }  
  
\def\preprint{\twocolumn\sloppy\flushbottom\parindent 2em  
        \leftmargini 2em\leftmarginv .5em\leftmarginvi .5em  
        \oddsidemargin -.5in    \evensidemargin -.5in  
        \columnsep .4in \footheight 0pt  
        \textwidth 10.in        \topmargin  -.4in  
        \headheight 12pt \topskip .4in  
        \textheight 6.9in \footskip 0pt  
        \def\@oddhead{\thepage\hfil\addtocounter{page}{1}\thepage}  
        \let\@evenhead\@oddhead \def\@oddfoot{} \def\@evenfoot{} }  
  
\def\numberbysection{\@addtoreset{equation}{section}  
        \def\theequation{\thesection.\arabic{equation}}}  
  
\def\underline#1{\relax\ifmmode\@@underline#1\else  
        $\@@underline{\hbox{#1}}$\relax\fi}

\def\titlepage{\@restonecolfalse\if@twocolumn\@restonecoltrue  
\onecolumn  
     \else \newpage \fi \thispagestyle{empty}\c@page\z@  
        \def\thefootnote{\fnsymbol{footnote}} }  
  
\def\endtitlepage{\if@restonecol\twocolumn \else \newpage \fi  
        \def\thefootnote{\arabic{footnote}}  
        \setcounter{footnote}{0}}  
  
\catcode`@=12  
\relax  
  
\def\figcap{\section*{Figure Captions\markboth  
        {FIGURECAPTIONS}{FIGURECAPTIONS}}\list  
        {Figure \arabic{enumi}:\hfill}{\settowidth\labelwidth{Figure  
999:}  
        \leftmargin\labelwidth  
        \advance\leftmargin\labelsep\usecounter{enumi}}}  
 \relax  
\def\tablecap{\section*{Table Captions\markboth  
        {TABLECAPTIONS}{TABLECAPTIONS}}\list  
        {Table \arabic{enumi}:\hfill}{\settowidth\labelwidth{Table  
999:}  
        \leftmargin\labelwidth  
        \advance\leftmargin\labelsep\usecounter{enumi}}}  
 \relax  
\def\reflist{\section*{References\markboth  
        {REFLIST}{REFLIST}}\list  
        {[\arabic{enumi}]\hfill}{\settowidth\labelwidth{[999]}  
        \leftmargin\labelwidth  
        \advance\leftmargin\labelsep\usecounter{enumi}}}  
 \relax  
\makeatletter  
\newcounter{pubctr}  
\def\publist{\@ifnextchar[{\@publist}{\@@publist}}  
\def\@publist[#1]{\list  
        {[\arabic{pubctr}]\hfill}{\settowidth\labelwidth{[999]}  
        \leftmargin\labelwidth  
        \advance\leftmargin\labelsep  
        \@nmbrlisttrue\def\@listctr{pubctr}  
        \setcounter{pubctr}{#1}\addtocounter{pubctr}{-1}}}  
\def\@@publist{\list  
        {[\arabic{pubctr}]\hfill}{\settowidth\labelwidth{[999]}  
        \leftmargin\labelwidth  
        \advance\leftmargin\labelsep  
        \@nmbrlisttrue\def\@listctr{pubctr}}}  
 \relax  
\makeatother  
\newskip\humongous \humongous=0pt plus 1000pt minus 1000pt

\newif\ifdtup

\font\Scbig=cmss10 scaled\magstep1  
\font\Scscr=cmss8 scaled\magstep1  
\font\Scscrscr=cmss8  
\newfam\Scfam  
\textfont\Scfam=\Scbig  
\scriptfont\Scfam=\Scscr  
\scriptscriptfont\Scfam=\Scscrscr

\relax  
\hybrid  
\def\lvm{\leavevmode\hbox to\parindent{\hfill}}  
  
\def\thefootnote{\arabic{footnote}}  
\def\BE{\begin{equation}}  
\def\EE{\end{equation}}  
\def\BA{\begin{eqnarray}}  
\def\EA{\end{eqnarray}}

\def\tt{\bar\tau}  
  
\def\lvm{\leavevmode\hbox to\parindent{\hfill}}  
\def\bar{\overline}

\def\BE{\begin{equation}}  
\def\EE{\end{equation} \vskip 0.30\baselineskip}  
\def\BA{\begin{array}}  
\def\EA{\end{array}}  
  
\def\noi{\noindent}  
  
\def\frac#1#2{{\textstyle{{#1}\over{#2}}}}

\newif\ifold \oldtrue   
\let\ssection=\section  
\def\section{\setcounter{equation}{0}\ssection}  
  
\begin{document}  
\renewcommand{\theequation}{\arabic{equation}}  
\newcommand{\beq}{\begin{equation}}  
\newcommand{\eeq}[1]{\label{#1}\end{equation}}  
\newcommand{\ber}{\begin{eqnarray}}  
\newcommand{\eer}[1]{\label{#1}\end{eqnarray}}  
\begin{titlepage}  
\begin{center}  
   
\hfill physics/0308078  
\vskip .4in  
  
{\large \bf Brane Worlds, 
the Subanthropic Principle and the Undetectability Conjecture} 
\vskip 0.9cm  
    
{\bf Beatriz Gato-Rivera} 
\vskip 1in    
  
  
\end{center}  
  
\begin{center} {\bf ABSTRACT } \end{center}  
\begin{quotation}  
In the recent article `Conflict between anthropic reasoning and observation'  
(gr-qc/0303070) Ken D. Olum, using some inflation-based ideas and the anthropic  
premise that we should be typical among all intelligent observers in the 
Universe, arrives at the puzzling conclusion that `we should find ourselves in  
a large civilization (of galactic size) where most observers should be, while in  
fact we do not'. In this note we discuss the intriguing possibility whether we 
could be in fact immersed in a large civilization without being aware of it. Our 
conclusion is that this possibility cannot be ruled out provided two conditions 
are met, that we call the Subanthropic Principle and the Undetectability 
Conjecture. The Subanthropic Principle states that we are not typical among the 
intelligent observers from the Universe. Typical civilizations of typical 
galaxies would be hundreds of thousands, or millions, of years more evolved 
than ours and, consequently, typical intelligent observers would be orders of 
magnitude more intelligent than us. The Undetectability Conjecture states that, 
generically, all advanced civilizations camouflage their planets for security 
reasons, so that no signal of civilization can be detected by external 
observers, who would only obtain distorted data for disuasion purposes. These 
conditions predict also a low probability of success for the SETI project. We 
also argue that it is brane worlds, and not inflation, what dramatically could 
aggravate the `missing-alien' problem pointed out first in the fifties by 
Enrico Fermi.

\end{quotation}  
\vskip 0.2cm  
  
August 2003   
\end{titlepage}  

\hfill{\ \ \ } 
\vskip 5cm
\hfill {\it To the memory of Giordano Bruno \ \ \ \ }
\newpage

\def\baselinestretch{1.2}  
\baselineskip 17 pt  

\noindent 
{\it Innumerable suns exist; innumerable earths revolve} \\
{\it around these suns in a manner similar to the way } \\
{\it the seven planets revolve around our sun.} \\
{\it Living beings inhabit these worlds.} 
 
Giordano Bruno, 1584 
 
\vskip 1cm 
 
\section{Preliminaries}\lvm  
  
Do mountain gorillas know that their `civilization' is embedded in a larger  
`civilization' corresponding to a much more evolved and intelligent species  
than themselves? Do they know that they are a protected species   
inhabitating a natural reserve in a country inside the African continent of  
planet Earth? The answer to these questions is certainly no, they do not   
know anything about our social structure, our countries, borders, religions,  
politics, ..... nor even about our villages and cities, except perhaps for  
those individuals living in a zoo, or adopted as pets.  
  
In the same way, the human civilization of planet Earth could be immersed  
in a much larger civilization unknowingly, corresponding to much more  
evolved and intelligent species than ourselves. After all, the sun is only 
a young star among thousands of millions of much  
older stars in our galaxy and the possible existence  
of such advanced civilizations is only a question of biological evolution  
doing its job, slowly but relentlessly through the millennia\footnote{In
addition, as the civilizations would reach some mastery in the field of
genetic engineering, the general tendency would be to `improve' themselves,
that is their own species, among many others, giving rise to an acceleration 
of the biological evolution at unimaginable rates. We thank several 
readers, especially Jim Bogan, for this important suggestion.}. If this  
happens to be the case it is quite sensible to assume that these individuals  
regard our planet as a natural reserve, full of animal and vegetal species,  
the Solar System being nothing but a small `province' inside their vast   
territory.  
 
In this situation, the answer to the usual remark `if there are advanced   
extraterrestrials around, why they do not contact us openly   
and officialy and teach us their science and technology?'   
seems obvious. Would any country in this planet send  
an official delegation to the mountain gorilla territory to introduce  
themselves `openly and officially' to the gorilla authorities? Would they  
shake hands, make agreements and exchange signatures with the  
dominant males? About teaching us their science and technology, who would 
volunteer to teach physics, mathematics and engineering to a bunch of  
gorillas? In addition one has to take into account the limits of the brain   
capabilities, independently of the culture or education. For example,  
let us ask ourselves how many bananas would be necessary for   
the most intelligent gorillas  
to understand the Maxwell equations of electromagnetism (even if they  
watch TV or listen to the radio). In the same way we may wonder how  
many sandwiches, potato chips or cigarretes would be necessary  
for the most intelligent among our scientists to understand the key  
scientific and technological results of a much more advanced 
civilization. Our intellectual  
faculties and habilities are limited by our brain capabilities that are  
by no means infinite. Therefore it is most natural and sensible to   
assume that there may exist important key scientific and 
technological concepts and  
results whose understanding is completely beyond the brain capabilities   
of our species, but is within reach of much more evolved and sophisticated  
brains corresponding to much more advanced civilizations.  
   
The motivation for this idea has been the recent article `Conflict between  
anthropic reasoning and observation' by Ken D. Olum \cite{Olum}. In this  
article the author presents some computations regarding the probabilities  
that typical intelligent observers belong to a large (galactic size)  
civilization at the present time. The underlying idea  
is that in the observable Universe, because of the existence  
of thousands of billions of stars older than the Sun, there must be huge 
civilizations much older than ours which could have spread widely through 
the Universe. (Although not mentioned in \cite{Olum}, Enrico Fermi was 
probably the first scientist to consider similar arguments, in the fifties,
leading to some `missing-alien' problem or paradox\footnote{We thank  
Juan Luis Ma\~nes and Cumrun Vafa for pointing out this fact to us.}.)  
In particular, using the  
assumption of an infinite Universe, like in models of eternal inflation,  
and doing some conservative computations, Olum predicts 
that `{\it all but one individual in $10^8$ belongs to  
a large civilization}'. Then he invokes the anthropic premise that we are  
typical individuals and, as a result, he predicts that there is a probability  
of $10^8$ versus 1 that we belong to a large civilization.  
Dropping the infinite Universe assumption,  
but keeping still inflation, the author claims that the predictions  
are not very different than for the previous case. After analysing several  
possibilities of where the problem might lie the author concludes:  
`{\it A straightforward application of anthropic reasoning and reasonable  
assumptions about the capabilities of other civilizations predict that we  
should be part of a large civilization spanning our galaxy. Although the  
precise confidence to put in such a prediction depends on one's  
assumptions, it is clearly very high. Nevertheless, we do not belong  
to such a civilization. Thus something should be amiss....... but then  
what other mistakes are we making.....?}'  
  
In this note we present what we think is the simplest possible solution 
to Olum's and Fermi's `missing-alien' problems and paradoxes. As we will 
discuss in detail, we could well be part of a large civilization  
spanning our galaxy (or a large region of it) without being aware of it.      
Therefore one obviously natural solution is that we do belong  
to a large, very advanced civilization, but we are not `citizens' 
of it because of our primitive low status. The two major mistakes of  
Olum's, therefore, would have been to assume: first, that we are typical 
intelligent observers, and, second, that to belong to a civilization  
implies to be a citizen of it.  
 
Besides, Olum's arguments implying that inflation must {\it necessarily}  
aggravate the (very serious) `missing-alien' problem do not seem very 
convincing and it is some brane world scenarios \cite{B-W}, in our opinion, 
what could in fact aggravate dramatically this problem. The reason is  
the following. If there exist thousands, or millions, of parallel universes  
separated from ours through extra-dimensions, it would be natural then to 
expect that some proportion of these universes would have the same laws of 
physics as ours (presumably half of these would be of matter  
and the other half of anti-matter), and many of the corresponding advanced 
civilizations would master the techniques to travel or `jump' through 
(at least some of) the extra dimensions. This opens up enormous possibilities 
regarding the expansion of advanced civilizations simultaneously through 
several parallel universes with the same laws of physics, resulting in 
multidimensional empires. It could even happen that the expansion to 
other parallel galaxies through extra dimensions could be easier, 
with lower cost, than the expansion inside one's own galaxy\footnote{The 
first scientists to consider extra-dimensions and parallel universes were 
probably Maxwell and Faraday in the 19th century. Outside the scientific realm 
this idea is many thousands of years old. At present we are still in a very 
premature phase in the study of brane worlds and we do not know whether these 
ideas are in fact realistic. Cumrun Vafa thinks that the fact that we do 
not see aliens around could be the first proof of the existence of brane 
worlds: all advanced aliens would have emigrated to better parallel 
universes (our Universe has zero measure) \cite{CuVa}.}. 

In many other universes, however, the laws of physics would be different,
corresponding perhaps to different vacua of the `would be' ultimate
Theory of Everything,  
resulting probably in `shadow matter' universes with respect to ours.  
This means that shadow matter would only interact with our matter 
gravitationally, in the case it would be brought to our Universe  
using appropiate technology. This does not mean, however, that the shadow 
universes would be necessarily empty of intelligent beings. 
If some of them had advanced civilizations, some of their 
individuals could even `jump' to our Universe, but not for colonization 
purposes since they would not even see our planets and stars, which
they would pass through almost unaware (they would only notice the
gravitational pull towards their centers). And the other way around,
we could neither see, nor talk to, the shadow visitors, although they could 
perhaps try to communicate with the `would be' intelligent beings of our 
Universe, through gravitational waves for example. Regarding anti-matter 
universes, the intelligent anti-observers would not send colonizers  
either\footnote{although they could send unwanted anti-prisoners,  
their arrival being known as gamma-ray bursts.}.  
 
We must also point out that in \cite{Olum} there is a continuous,  
repeatedly use of the concept `intelligent observer' without a 
definition of its meaning. This fact makes difficult to follow Olum's 
arguments and computations properly. For example, do the Cro-Magnon and 
Neandertal mankinds qualify as civilizations of intelligent observers? 
How about the very primitive human beings living nowadays in some forests? 
Do they qualify? Do they belong to `the civilization' of planet Earth even 
if they know very little about it?  
  
For the discussion in next sections we will  
use the following intuitive definitions: 
   
{\it Primitive civilizations}: Are those civilizations that have a 
remarkable use of technology in everyday life but are uncapable of 
leaving their planets to colonize other ones in different stellar 
systems. Their scientific knowledge can have many degrees, 
ranging from zero to remarkable high levels. In our planet it seems 
that only the groups of human beings from, approximately, the last 
20.000 years qualify as primitive civilizations, corresponding to what
anthropologists call the Modern Man, not so the groups of the various
versions of the Early Man, who would only qualify as {\it very 
primitive civilizations}. We call the individuals of the primitive 
civilizations {\it primitive intelligent observers}. 
   
{\it Advanced civilizations}: Are those civilizations technologically 
able to colonize other planets from different stellar systems, 
ranging from a few planets until thousands of them or more in the 
case of {\it very advanced civilizations}. Depending on their 
technological level they could even travel through extra dimensions 
(if they exist) and they could visit and colonize planets located in 
`would be' nearby galaxies in parallel universes. We call the individuals 
of these civilizations {\it advanced} and 
{\it very advanced intelligent observers}, respectively.

\section{The Main Ideas}\lvm  
 
Let us discuss in detail the possibility that our small terrestrial
civilization is embedded in a large civilization unknowingly. 
This will lead very naturally to the the proposal 
of two major ideas that we call the `Subanthropic Principle' and  
the `Undetectability Conjecture'. 
     
To start let us come back to the main argument. In our galaxy there 
are thousands of millions of stars much older than the Sun, many of
them thousands of millions of years older, in fact. 
Therefore, it seems most natural to expect,  
without the need to invoke inflation, that in a reasonable   
amount of stellar systems technological civilizations   
should have appeared, and a fraction of them (even tiny) should have  
survived enough to spread to, at least, large regions of the galaxy.  
It is then very remarkable the fact that the Solar System has never been   
encountered or colonized by any advanced civilization,..... or has it?   
  
In our opinion there is an important flaw in Olum's (implicit) assumptions  
about the relations between the different civilizations put into contact  
in the process of expansion. Although he does not mention this crucial issue,  
one gets the impression that he believes that the more advanced civilizations   
will push the less advanced ones to their own level in order to integrate  
them, or rather they will exploit, damage, or annihilate them in order to  
conquer the planet, in the case of aggressive colonizers. We fully agree  
that aggressive advanced civilizations will exploit/damage/annihilate  
the less advanced civilizations as much as it is convenient for them.  
In the case of non-aggressive advanced civilizations, however, the   
possibility that they will integrate the less advanced ones only makes  
sense if those ones are not that inferior. That is, if the gap between  
the two civilizations is not very big, then it is realistic to expect  
that the superior civilization will push the inferior one to their  
own level, to some extent at least. In some cases, however, the   
non-aggressive advanced civilizations will encounter planets with  
primitive or very primitive civilizations, with an enormous gap   
(technologically, scientifically and genetically) between them. 
In particular, the differences between their brain capabilities  
and those of the primitive individuals could be pathetic. In these  
circumstances, it is completely unrealistic and naive to expect that  
the advanced individuals will try to integrate the primitive ones into  
their own civilizations. They rather will behave `ecologically' towards  
them, treating them as sort of `protected species' and not interfering  
(or only very discretely) with their natural evolution.  
  
With this insight it is now much easier to accept the possibility that  
the Solar System could have been encountered or colonized many   
thousands, or even millions, years ago by, at least, one non-aggressive  
advanced civilization, who treated and still treat our planet as some   
protected natural reserve. As a matter of fact, they could have even 
brought many plants and animals to planet Earth, including our ancestors, 
presumably to improve their life conditions (they could have been 
in danger of extinction in their own planet, for example)\footnote{One
expected activity of advanced civilizations would be the dissemination 
of life on `promising' planets, in the same way that we plant trees in 
appropriate environments. In the case life started in this way on planet 
Earth, then all terrestrial living beings would have common building blocks 
of DNA with the living beings of thousands of other planets which would
have undergone similar insemination procedures with the same bacteria.
It is conceivable therefore that, under these circumstances, plants and
animals could have been brought to Earth whose extraterrestrial origin 
would be impossible to detect by any biologist or geneticist.}.  
Perhaps the Solar System has been visited by aggressive colonizers, as  
well as non-aggressive ones, resulting in battles or just pacific  
negotiations between them. Perhaps the aggressive losers will come back 
in the future, to try again. 
 
This view about ourselves, a small primitive civilization immersed in 
a large, advanced civilization, leads straightforwardly to the realization 
that we could find ourselves not among the typical intelligent observers of 
our galaxy, but only among a small proportion of primitive intelligent  
observers instead, completely ignorant of their low status. The typical
intelligent observers would be the citizens of the advanced and very
advanced civilizations who would `own' the galaxy. But our 
galaxy is just one typical galaxy from our observable Universe. This 
leads very naturally to our first proposal: 
  
{\it The Subanthropic Principle}: We are not typical among the intelligent  
observers from the Universe. Typical civilizations of typical galaxies would  
be hundreds of thousands, or millions, of years more evolved than ours and,  
consequently, typical intelligent observers would be orders of magnitude more  
intelligent than us.  

Observe that the Subanthropic Principle is almost equivalent to the 
proposal that, at present, all typical galaxies of the Universe are 
already colonized (or large regions of them) by advanced, or very
advanced, civilizations, a small proportion of their individuals 
belonging to primitive subcivilizations, like ours.  
Whether the primitive subcivilizations know or ignore their 
low status will, most likely, depend on the ethical standards  
of the advanced civilization in which they are immersed. If the standards 
are low, the individuals of the primitive subcivilizations will be 
surely abused in many ways, in the same way that in our civilization 
large groups of human beings abuse other human beings in weaker positions, 
as well as animals in general. Therefore, in this case the primitive  
individuals will be painfully aware of their low status. If the 
ethical standards of the advanced individuals is high instead, 
then very probably they will respect the natural evolution (biological, 
social, cultural) of the primitive subcivilizations, treating them
`ecologically' as some kind of protected species. In this case, that we 
think could well describe the situation of the terrestrial civilization, 
the primitive individuals would be completely unaware of the existence 
of the large advanced civilization in which they are immersed.  
  
Now there is an important remark: if the Solar System is part of the  
territory of an advanced civilization, why we do not detect any signal  
of civilization in any of the solid planets and large satellites in it?  
It would be most natural if they had built bases all along the Solar 
System (including underground and submarine bases in planet Earth)
and maybe some colonies here and there on, or beneath, the surface  
of some solid planets and large satellites  
(this is exactly what we plan to do in the future ourselves!).  
The simplest answer would be that they do not find the Solar System  
attractive enough to live themselves and, as a consequence, they have  
only a few tiny bases difficult to detect. However, independently of  
whether or not they find the Solar System attractive to build colonies,  
we believe that all advanced civilizations must be necessarily aware of  
the existence of {\it aggressive} advanced civilizations and, as a result, 
they should have developed very sophisticated camouflage systems, so that   
no signals of civilization can be detected by any external observers 
(neither by their space probes). Probably, in many cases they even  
manipulate and distort the global data of their planets (temperature, air  
composition, etc.), to fool external observers for disuasion  
purposes\footnote{It may sound strange that advanced civilizations would 
need to protect themselves against aggressors. However, there is not a 
single proof or indication that the ethical development of a civilization 
or an individual, grows in parallel with their level of material 
well-being or with their technological and scientific development.
One may also argue that advanced aggressive civilizations must annihilate
themselves, what seems a sensible guess. The crucial issue, however, is
not whether they will annihilate themselves but {\it how much damage}
they can produce to other civilizations (primitive as well as advanced)
before they annihilate themselves.}.  
This is the content of our second proposal: 

\newpage
 
{\it The Undetectability Conjecture}: Generically, all advanced enough  
civilizations camouflage their planets for security reasons, so that  
no signal of civilization can be detected by external observers, who  
would only obtain distorted data for disuasion purposes.  
 
Observe that, if this conjecture turns out to be true, then we cannot 
be sure whether the terrestrial civilization is the unique 
civilization inhabitating the Solar System, as we firmly believe
(this statement is independent, in fact, of whether or not our
civilization is embedded in a large advanced civilization, we
only needed to have `advanced neighbors'). 
In fact, the inconsistency in the scientific reasoning used in the 
astronomical observations of planets and satellites is remarkable. 
One uses as input the non-proved assumption that, at the 
sources, there are no intelligent beings manipulating the data that we 
receive, and then one concludes that there is no signal of intelligent 
life, {\it as the data prove}. But this assumption could turn out to 
be wrong. The right claim would be in this case that there is no signal 
of {\it primitive civilizations}, like ours, who would allow themselves 
to be detected by external observers, but {\it nothing} can be 
said about the possibility of {\it advanced civilizations}, capable 
to fool our telescopes, detectors and space probes, and who 
would not allow themselves to be detected.  
   
Finally, we must mention that the first scholar, at least in  
western history, who suggested that many stars out there 
could have planets similar to ours: with plants, animals, people, etc., 
was Giordano Bruno, in the 16th century. He claimed that the Sun was 
only one star among the many thousands, and therefore, like the Sun, 
many other stars would also have planets around and living beings 
inhabitating them \cite{Bruno}. To appreciate the genius of Giordano 
Bruno one has to take into account that he lived at the time 
when more than 99\%  of the intellectuals believed that the Earth was 
the center of the Universe, and a few others, like Copernico 
and Galileo, believed that it was the Sun, instead, the center of  
the Universe, the stars being some bright heavenly bodies of unknown 
nature\footnote{For these and other ideas Giordano Bruno was imprisoned 
eight years and finally burned at the stake in Rome, in piazza Campo di 
Fiori, the 17th February 1600. The catholic church, which some years ago 
apologized for the treatment given to Galileo, has never apologized, 
however, for the treatment received by Giordano Bruno.}. Nowadays we
know that the Universe has no center and that our planet is only a tiny
particle of dust in its inmensity. In spite of this, for many human
beings the Earth is still the center of the Universe, the chosen planet
inhabited by the most perfect and intelligent beings all over the 
Universe: the Crown of the Creation. (There are even regular scientists
and `intellectuals' who wonder whether the whole Universe was created 
just for us, terrestrial human beings, to exist!). 

\newpage
  
\section{Conclusions and Final Remarks}  
 
We have discussed the possibility that our civilization could 
be embedded in a large advanced civilization spanning (at least) 
a large region of our galaxy. This should be expected, in fact, 
since in our galaxy there are many thousands of millions of  
stars much older that the Sun. Using two simple and natural  
assumptions we see that this possibility cannot be ruled out. 
 
The first assumption explains why the members or citizens 
of the large civilization would not interact and socialize with us 
(openly and officially, at least). The reason would be that 
we do not qualify as full members, neither as associates, nor 
to be in the queue for applications, although we perhaps 
qualify as pets or `friends'. This situation we generalize, 
taking into account that we live in a typical galaxy, resulting in the 
Subanthropic Principle that states that we are not typical among the 
intelligent observers from the Universe, but much below the standards. 
 
The second assumption, that we call the Undetectability Conjecture, 
explains why we do not detect any signals of this large 
civilization in which we would be immersed. The reason would 
be that, generically, all advanced civilizations are undetectable 
for security reasons, due to the existence of {\it aggressive} advanced 
civilizations. In any case, why would advanced civilizations 
allow alien civilizations to watch their cities,  
laboratories, military installations, etc., when they could fool 
them very easily instead? 
 
The Subanthropic Principle is almost equivalent to the proposal that all 
the typical galaxies of the Universe are already colonized (or at least 
large parts of them) by advanced, or very advanced, civilizations, which 
is a most natural guess taking into account that many 
thousands of millions of stars which populate the typical galaxies
are thousands of millions of years older than the Sun. 
In these large advanced civilizations there would always exist, 
generically, a small percentage of individuals which belong to primitive 
subcivilizations. If the ethical standards of the advanced individuals 
are low, then the primitive individuals will be abused in many ways 
(maybe even annihilated). If the ethical standards of 
the advanced individuals are high instead, then probably they 
will treat the primitive individuals in an ecological way; that is, 
like a protected species living in a natural reserve. In this case, 
which could well describe the situation of our civilization, 
most of the primitive individuals would completely ignore 
the existence of the advanced civilization in which they are immersed.
 
We have also argued that the idea of brane worlds, although still 
in very premature stage, could in fact aggravate enormously the 
`missing-alien' problem, pointed out first by Enrico Fermi as was 
mentioned before. The reason is 
that, if other parallel universes exist with the same laws of physics than 
ours, it could happen that advanced civilizations could be technically 
able to `jump' through the extra dimensions to our galaxy for expansion 
and colonization purposes. As a result, it could even happen that 
the `owners' of the Solar System (if they exist) had come from another 
universe and had created a huge multidimensional empire, with large 
pieces of territory in several `parallel' galaxies. It could also happen 
that advanced civilizations would find more efficient 
(cheaper, energetically preferable) to expand along extra 
dimensions than inside their own galaxy. 
 
Finally, in the Appendix we discuss the issue of possible contacts and
interactions between advanced civilizations or individuals and primitive 
civilizations or individuals. It seems very unlikely that non-aggressive
advanced civilizations would `introduce themselves' to any primitive
civilization. Nevertheless, we have identified three major causes or 
reasons which could motivate individuals of advanced civilizations to seek
interactions or relationships with primitive individuals: scientific
purposes, entertainment purposes and criminal purposes. We also point out 
that the Subanthropic Principle and the Undetectability Conjecture predict
a rather low probability of success for the SETI project, the reason 
being the low percentage of technological civilizations 
susceptible to be detected (the period of detectability of an
average civilization could last less than 500 years). 

\vskip .8cm

\centerline{\large \bf Appendix}

\vskip .4cm

In what follows we will discuss the possible sources of contacts and
interactions between advanced civilizations or individuals and primitive 
civilizations or individuals. As we argued in the preliminaries, it is very
unlikely that a non-aggressive advanced civilization would contact any
primitive civilization `openly and officially' (at least until the latter
reaches a remarkable degree of development that our civilization has not
reached yet). Aggressive advanced civilizations, however, would `introduce
themselves' before, after, or during the strike, for their own convenience. 
(The fact that our civilization has never been attacked by aggressive 
aliens, as far as history knows, could be indeed a clue that we belong 
to a non-aggressive advanced civilization which protects planet Earth, 
as part of its territory).

If we now consider possible contacts and relationships between individuals 
of advanced civilizations and primitive individuals, rather than between 
their civilizations, many more possibilities appear. Trying to identify 
which advanced individuals could seek interactions or relationships with 
primitive individuals, and for which reasons, leads us to distinguish 
three main sources of contacts: 
 
1) Scientific research by regular scientists related to life-sciences, 
such as biologists, medical researchers, anthropologists, sociologists, 
psychologists, etc. Whether or not the corresponding research activities 
could damage the primitive individuals (physically or mentally), would 
depend on the legal regulations of the advanced civilizations regarding 
ethical treatment towards individuals of primitive civilizations. 
 
2) Entertainment, affection, etc. That is, one reason for an individual 
of an advanced civilization to establish contact with primitive 
individuals could be simply to have fun and relax. The advanced 
individual could have, with respect to the primitive individuals, 
the kind of feelings that push us to interact and play with cats 
and dogs and many other species. In addition, if in our planet 
there are millions of cat-lovers and millions of dog-lovers, and there 
are even snake, pig, .... and gorilla-lovers, it is most natural 
to expect that there may exist some primitive-individual-lovers,
in particular terrestrial-human-lovers, among 
advanced aliens. Why not? This would apply 
specially to those advanced individuals who must spend long periods of
time working in primitive planets, living in underground or submarine 
boring bases, which surely would exist in our planet in the case that 
our civilization is embedded in a large civilization (the workers in 
the bases being the `guards' or militars that take care of the planet).
 
3) Criminal purposes of all kinds, including activities by regular 
scientists which would be 
forbidden by their legal ethical regulations. We can 
imagine dozens of different criminal purposes for which the primitive 
individuals could be kidnapped, tortured and even killed, including 
abject topics such as `high gastronomy' and sadist games. To be 
realistic, one only has to think of the treatment that some cruel human 
beings give to their victims, be either other human beings (often children)
or animals. The point is that the ethical level of an individual, or
a civilization, does not necessarily grow in parallel with their
technological and scientific achievements, or with their level of material
well-being. In the case that our civilization is embedded in a large
civilization, one of the tasks of the `guards' living in the bases would
be, undoubtely, to chase away the human-hunters and other outlaws. 
 
\vskip 0.1cm 
   
Regarding the SETI - search for extraterrestrial intelligence - 
project, if the Undetectability Conjecture turns out to be true, 
then SETI turns into SETPI: search for extraterrestrial {\it primitive} 
intelligence. The reason is that, in this case, 
only primitive civilizations could be detected by external observers. 
On the other hand, if the Subanthropic Principle is correct, then the 
primitive civilizations would be very scarce compared to the total amount
of technological civilizations, and even more scarce would be the ones
with an appropriate technological level to produce electromagnetic 
emissions to be detected by distant civilizations. (Observe that the 
period of `detectability' of an average civilization could last less 
than 500 years). Therefore the probability 
for one primitive civilization to detect another one would be very small. 
For these reasons the Subanthropic Principle and the Undetectability 
Conjecture predict a rather low probability of succes for the SETI 
project\footnote{Maybe the SETI-experts 
should join the competition for collaboration: the `human antennas' or  
`alternative-SETI'-experts, who claim to have long-term, 
well established contacts with alien partners, some of them
`sitting at home' in their own planets and some others living 
temporarily in our planet in underground or submarine basis.
The guiness record in such relations is probably held by the spanish 
`grupo Aztlan', which for around {\it twenty five years} 
gather once per week at night to establish (what they claim to be) 
telepathical communications with a group of sociologists of planet Apu,  
orbiting Alpha B Centauri, which are sitting at home, or in the office.}.

A last remark is that we have never done any investigation in the subject 
of alledged alien contacts. As a result we have essentially no opinion 
about the true or false alien nature of those
circulating through the media. Nevertheless, 
we believe that it must be an impossible task to identify true alien 
contacts (if they exist at all) just by reading the reports given to,
or written by, their terrestrial partners. The reason is that, for our 
intuition, the claims of civilizations much more advanced than us must 
{\it necessarily} sound ridiculous, hilarious, crazy, science fiction 
ideas. But the same would have happened if we had described our TV sets,
our planes, our microwave ovens, our computers, etc .....to people only 
100 years ago!. Let us also notice that many persons, including many
scientists, have a very deep rooted reluctance and aversion to accept 
the possibility of the existence of extraterrestrial species much 
more advanced and intelligent than us, who could even visit our planet.
We call this prejudice the `Crown of the Creation Syndrome' (CCS), for 
obvious reasons. Curiously, whereas many religious persons are not CCS 
sufferers, many atheists are (one reason could well be that they 
grew in very religious families which implanted in their children's 
minds strong impressions of the greatness and uniqueness of the human 
species)\footnote{In short, the CCS sufferers firmly believe, or secretely
hope, that nobody in the whole Universe can do what we cannot do, in
particular interstellar travel. Interestingly, they often harbour very
high expectations about the future capabilities and great achievements
awaiting our civilization. In spite of this, in their reasoning they lack
the ability to interchange `we' by `they', and the future by the past,
in reference to possible civilizations millions of years older than ours.
For example, many of them accept gladly big suggestions of the type `we will
travel to other planets and stars', `we will colonize the galaxy', etc.
and however, they cannot even listen to the suggestions that `they could
have travelled to other planets and stars (including the Solar System
and the Earth)', `they could have colonized the galaxy', etc., which
are received with sarcasm, disapproval and even anger.}.

\vskip .1cm

To finish, we would like to point out that the present status of the
Search for ExtraTerrestrial Intelligence could well be described by the 
popular american protest-song from the sixties that we transcribe next.

\vskip .20in

{\it {\ \ \ } Where have all the aliens gone?} 

{\it {\ \ \ } Long time passing....}

{\it {\ \ \ } Where have all the aliens gone?}

{\it {\ \ \ } Long time ago....}

{\it {\ \ \ } Where have all the aliens gone?}

{\it {\ \ \ } Could be hidden everywhere!}

{\it {\ \ \ } When will we ever learn?}

{\it {\ \ \ } When will we ever learn?}

{\ \ \ \ }(Repeat three times)

\vskip .4in

\noi
\centerline{\bf Acknowledgements}

\vskip .2in 

We are grateful to many readers of this article who have expressed their
appreciation for it. After it was published in Popular Physics and Space
Physics exactly one year ago, we have received an avalanche of questions,
comments, suggestions, observations,... and also a considerable amount
of information in the form of bibliography and webpages related to the 
matters discussed by us. Unfortunately, we have not had the opportunity 
to read more than a minimum piece of that information, for lack of time. 
Of special interest is the website www.ufoskeptic.org, by astrophysicist 
Bernard Haisch, which has been made for and by scientists with some interest
in the subject of alien civilizations. We are also grateful and indebted to 
our friend Mar\'\i a Teresa Fern\'andez Mart\'\i nez for many illuminating
conversations and for her help in preparing the spanish translation of the
article. We also thank Bill Parkyn for correcting several english 
typos, as well as for his interesting remarks. Finally, we appreciate the 
cartoon www.funny-city.com/cartoons/images/43.jpg  which has been sent to us.

\vskip .4in  
\noi
{\bf Note}
\vskip .2in 

The first version of this article appeared in August 2003. This date
remains on the cover of the present revised version, finished in August 
2004, because the article remains essentially the same, except for several
minor modifications: four footnotes (1, 5, 8 and 9), one last paragraph 
in the Appendix and a few small improvements in the redaction. We have 
translated this article to spanish after we received several requests 
to do so. The spanish version is also available at physics/0308078.
Finally, Popular Science is not officially supported
by the Spanish Scientific Research Council (CSIC) and should
be done therefore as a private activity. For this reason there is no
preprint number and affiliation data on the cover of this article.

\vskip .4in  
\noi
{\bf Biographical Note}
\vskip .2in 

The author, Beatriz Gato Rivera, was born in Madrid, Spain. She 
studied Physics in the Complutense University in Madrid, where
she also obtained the Ph.D. in Theoretical Physics in 1985. She
is specialist in Particle Physics and Mathematical Physics. After
spending three postdoctoral years at MIT (Masachussets Institute
of Technology) and another three years at CERN, she joined the
permanent research staff of the Spanish Scientific Research Council 
(CSIC). e-mail: t38@nikhef.nl 
 

\vskip .4in  
   
\begin{thebibliography}{9}  
\def\NPB{Nucl. Phys. B}  
\def\PLB{Phys. Lett. B}  
\def\MPLA{Mod. Phys. Lett. A}  
\def\IJMPA{Int. J. Mod. Phys. A}  
  
\bibitem{Olum} K. Olum, `Conflict between anthropic reasoning and  
observation', gr-qc/0303070.  
 
\bibitem{B-W} N. Arkani-Hamed, S. Dimopoulos and G.R. Dvali, 
Phys. Lett. B429, 263, 1998; Phys. Rev. D 59, 86004, 1999;
Phys. Today 55N2, 35, 2002.\\
L. Randall and R. Sundrum, Phys. Rev. Lett. 83, 4690, 1999.
 
\bibitem{CuVa} C. Vafa, private communication.

\bibitem{Bruno} G. Bruno, {\it On the Infinite Universe and Worlds}, 1584. 
    
\end{thebibliography}
\end{document}


\renewcommand{\theequation}{\arabic{equation}}  
\newcommand{\beq}{\begin{equation}}  
\newcommand{\eeq}[1]{\label{#1}\end{equation}}  
\newcommand{\ber}{\begin{eqnarray}}  
\newcommand{\eer}[1]{\label{#1}\end{eqnarray}}

\begin{titlepage}  
\begin{center}

\hfill physics/0308078  
\vskip .3in  
  
{\large \bf Universos Branas,
el Principio Subantr\'opico y la Conjetura de Indetectabilidad} 

\vskip 0.7cm

{\bf Beatriz Gato-Rivera}  

\vskip 0.5cm  
  
\end{center}  
  
\begin{center} {\bf ABSTRACT } \end{center}  
\begin{quotation}  
En el reciente art\'\i culo `Conflicto entre razonamiento antr\'opico y observaci\'on'  
(gr-qc/0303070) Ken D. Olum, bas\'andose en algunas ideas sobre la inflaci\'on cosmol\'ogica y en la premisa antr\'opica de que tenemos que ser t\'\i picos entre
los observadores inteligentes del Universo, llega a la sorprendente conclusi\'on
de que `deber\'\i amos encontrarnos en una civilizacion grande (de tama\~no
gal\'actico) al igual que la mayor\'\i a de los observadores, mientras que de 
hecho no lo estamos'.
En este art\'\i culo discutimos la posibilidad de que nuestra 
civilizaci\'on est\'e de hecho inmersa en una civilizaci\'on grande, sin que
seamos conscientes de ello. Nuestra conclusi\'on es que esta posibilidad no puede
descartarse si se cumplen dos condiciones, que llamamos el Principio Subantr\'opico
y la Conjetura de Indetectabilidad. El Principio Subantr\'opico establece que
no somos t\'\i picos entre los observadores inteligentes del Universo. Las 
civilizaciones t\'\i picas de las galaxias t\'\i picas estar\'\i an cientos de miles,
o millones, de a\~nos m\'as evolucionadas que la nuestra y, consecuentemente, los
observadores inteligentes t\'\i picos ser\'\i an \'ordenes de magnitud m\'as 
inteligentes que nosotros. La Conjetura de Indetectabilidad establece que,
gen\'ericamente, todas las civilizaciones avanzadas camuflan sus planetas por
razones de seguridad, de manera que los observadores externos no puedan detectar
se\~nal alguna de civilizaci\'on y s\'olo obtengan datos distorsionados con
el prop\'osito de disuasi\'on. Estas condiciones predicen tambi\'en una baja
probabilidad de \'exito para el proyecto SETI. Tambi\'en argumentamos que
son los universos branas, y no la inflaci\'on cosmol\'ogica, lo que podr\'\i a
agravar dram\'aticamente el problema del `alien\'\i gena ausente', se\~nalado
por vez primera en los a\~nos cincuenta por Enrico Fermi.

\end{quotation}  
\vskip 0.2cm  
  
Agosto 2003 

(Traducci\'on al castellano del original ingl\'es en Agosto 2004)  
\end{titlepage}  

\hfill{\ \ \ } 
\vskip 5cm
\hfill {\it A la memoria de Giordano Bruno \ \ \ \ }
\newpage

\def\baselinestretch{1.2}  
\baselineskip 17 pt  

\noindent 
{\it Innumerables soles existen; innumerables tierras giran alrededor} \\
{\it de estos soles de manera similar a la forma en que los} \\
{\it siete planetas giran alrededor de nuestro sol.} \\
{\it Seres vivos habitan estos mundos.} 
 
Giordano Bruno, 1584 
 
\vskip 1cm 
 
\section{Preliminares}\lvm  

\que Saben los gorilas de monta\~na que su `civilizaci\'on' est\'a inmersa en
una `civilizaci\'on' m\'as grande correspondiente a una especie mucho 
m\'as evolucionada e inteligente que ellos mismos? \que Saben que son una
especie protegida que habita en una reserva natural en un pa\'\i s dentro
del continente africano del planeta Tierra? La respuesta a estas preguntas 
es ciertamente no, ellos no saben nada sobre nuestra estructura social,
nuestros pa\'\i ses, fronteras, religiones, pol\'\i tica,..... ni siquiera sobre
nuestros pueblos y ciudades, con la excepcion quiz\'as de aquellos individuos
que viven en zoos o han sido adoptados como animales de compa\~n\'\i a. 
 
De la misma manera, la civilizaci\'on humana del planeta Tierra podr\'\i a
estar inmersa en una civilizaci\'on mucho m\'as grande sin saberlo,
correspondiente a una especie mucho m\'as evolucionada e inteligente. 
Al fin y al cabo, el Sol es s\'olo
una estrella j\'oven entre miles de millones de estrellas mucho m\'as
antiguas en nuestra galaxia, y la posible existencia de tales
civilizaciones avanzadas es s\'olo cuesti\'on de la evoluci\'on 
biol\'ogica haciendo su trabajo, lenta pero implacablemente a trav\'es
de los milenios\footnote{Adem\'as, a medida que las civilizaciones fueran
alcanzando cierta destreza en el campo de la ingenier\'\i a gen\'etica, la
tendencia general ser\'\i a a `mejorar' la propia especie, entre otras, 
con lo que se acelerar\'\i a la evoluci\'on biol\'ogica a ritmos 
inimaginables. Damos las gracias a varios lectores, especialmente Jim Bogan,
por esta sugerencia tan 
importante.}. Si \'este resultara ser el caso, ser\'\i a bastante
razonable suponer que estos individuos considerar\'\i an nuestro planeta como
una reserva natural, llena de especies animales y vegetales, siendo el
Sistema Solar s\'olo una peque\~na provincia dentro de su vasto territorio.   

En esta situaci\'on, la respuesta a la observaci\'on usual `si hay
extraterrestres avanzados cerca de nosotros, \que por qu\'e no 
entran en contacto abierta y oficialmente y nos ense\~nan su ciencia 
y tecnolog\'\i a?', parece obvia. \que Enviar\'\i a alg\'un pa\'\i s
de este planeta una delegaci\'on oficial al territorio del gorila de
monta\~na para presentarse abierta y oficialmente a las `autoridades'
de los gorilas? \que Se estrechar\'\i an la mano, llegar\'\i an a acuerdos
e intercambiar\'\i an firmas con los machos dominantes? Respecto a
ense\~narnos su ciencia y tecnolog\'\i a, \que qui\'en se presentar\'\i a
voluntario para ense\~nar f\'\i sica, matem\'aticas e ingenier\'\i a
a un grupo de gorilas? Adem\'as hay que tener en cuenta los l\'\i mites
de las capacidades cerebrales, independientemente de la cultura o la
educaci\'on. Por ejemplo, vamos a preguntarnos cuantos pl\'atanos 
ser\'\i an necesarios para que los gorilas m\'as inteligentes pudieran
entender las ecuaciones de Maxwell del electromagnetismo (incluso aunque
vieran la televisi\'on o escuchasen la radio). De la misma manera podemos
preguntarnos cuantos sandwiches, patatas fritas o cigarrillos ser\'\i an
necesarios para que los m\'as inteligentes de entre nues\-tros cient\'\i ficos 
pudieran entender los resultados cient\'\i ficos y tecnol\'ogicos clave de 
una civilizaci\'on mucho m\'as avanzada. Nuestras facultades y habilidades
intelectuales est\'an limitdas por nuestras capacidades cerebrales, que
no son en absoluto infinitas. Por lo tanto, es de lo m\'as natural y
razonable suponer que puedan existir conceptos y resultados cient\'\i ficos 
y tecnol\'ogicos clave, importantes, cuya comprensi\'on quede 
totalmente m\'as all\'a de las capacidades cerebrales de nuestra especie,
pero est\'e al alcance de cerebros mucho m\'as evolucionados y
sofisticados correspondientes a civilizaciones mucho m\'as avanzadas. 
    
La motivaci\'on que subyace en esta idea ha sido el art\'\i culo 
reciente `Conflicto entre razonamiento antr\'opico y 
observaci\'on'\footnote{El art\'\i culo [1] est\'a escrito 
en ingl\'es y es nuestra la traducci\'on del t\'\i tulo
`Conflict between anthropic reasoning and observation'.} de 
Ken D. Olum [1]. En este art\'\i culo el autor presenta algunos 
c\'alculos respecto a las probabilidades de que los observadores
inteligentes t\'\i picos de nuestro Universo pertenezcan a 
civilizaciones grandes (de tama\~no gal\'actico) en el tiempo presente.
La idea es que en el Universo observable, debido a la existencia de 
billones\footnote{Aqu\'\i\ nos referimos a billones europeos $(10^{12})$ que son 
mil veces m\'as grandes que los billones ameri- canos $(10^9)$ a que nos
refer\'\i amos en la versi\'on original de este art\'\i culo en ingl\'es. 
En consecuencia la expresi\'on `thousands of billions' la hemos traducido 
como `billones'.} de estrellas mucho m\'as antiguas que el Sol, tienen que 
existir civilizaciones gigantescas mucho m\'as antiguas que la nuestra,
que podr\'\i an haberse expandido ampliamente a trav\'es del Universo.
(Aunque no se menciona en [1], Enrico Fermi fue probablemente el primer 
cient\'\i fico en considerar argumentos similares, en los a\~nos cincuenta, 
que conducen a lo que podr\'\i amos denominar el problema o la paradoja del 
`alien\'\i gena ausente', conocida en la literatura como `la Paradoja de 
Fermi'\footnote{Damos las gracias a Juan Luis Ma\~nes y a Cumrun Vafa por
esta informaci\'on.}.) En particular, haciendo la suposici\'on
de que el Universo es infinito, como en los modelos de inflaci\'on
cosmol\'ogica perpetua (`eternal inflation'), y haciendo 
algunos c\'alculos conservadores, Olum predice que `{\it all but one 
individual in $10^8$ belongs to a large civilization}'; es decir, que entre 
cien millones de individuos, todos excepto uno pertenecen a una
civilizaci\'on grande. Despu\'es invoca la premisa antr\'opica de que 
somos individuos t\'\i picos y, como resultado, predice que hay una
probabilidad de $10^8$ sobre $1$ de que nosotros pertenezcamos a una
civilizaci\'on grande. Prescindiendo de la suposici\'on de que el
Universo sea infinito, pero manteniendo a\'un la inflaci\'on, el
autor asegura que las predicciones no son muy diferentes de las del
caso previo. Despu\'es de analizar varias posibilidades de d\'onde puede
estar el problema, el autor concluye: `{\it Una aplicaci\'on directa del
razonamiento antr\'opico y suposiciones razonables sobre las capacidades
de otras civilizaciones predicen que nosotros deber\'\i amos formar
parte de una civilizaci\'on grande extendida por nuestra galaxia.
Aunque la fiabilidad precisa de una predicci\'on como \'esta 
depende de las suposiciones que uno haga, es claramente muy alta.
Sin embargo, nosotros no pertenecemos a una tal civilizaci\'on. As\'\i\
que algo debe de estar equivocado....... pero entonces, \que qu\'e otros 
errores estamos cometiendo......?}'  

En este art\'\i culo presentamos lo que, creemos, es la soluci\'on m\'as
simple posible a los problemas y paradojas de Olum y Fermi sobre los
`alien\'\i genas ausentes'. Como discutiremos en detalle, nosotros
podr\'\i amos muy bien formar parte de una civilizaci\'on grande 
extendi\'endose por toda nuestra galaxia (o una regi\'on amplia
de la misma) sin ser conscientes de ello. Por tanto, una soluci\'on
natural obvia ser\'\i a que pertenecemos a una civilizaci\'on grande
muy avanzada, pero no somos `ciudadanos' de \'esta debido a nuestro bajo 
nivel primitivo. Los dos errores principales de Olum, pues, habr\'\i an
sido el suponer: primero, que somos observadores inteligentes 
t\'\i picos, y segundo, que pertenecer a una civilizaci\'on implica
ser ciudadano de la misma. 

Aparte, los argumentos de Olum respecto a que la inflaci\'on tiene
necesariamente que agravar el problema (muy serio de por s\'\i ) del
`alien\'\i gena ausente', no nos parecen muy convincentes, y ser\'\i an
sin embargo algunos modelos de `universos branas' [2] lo que en 
nuestra opini\'on podr\'\i a agravar dram\'aticamente este problema. 
La raz\'on es la siguiente. Si existieran miles, o millones, de universos
paralelos separados del nuestro a trav\'es de dimensiones extra, entonces 
ser\'\i a natural esperar que una proporci\'on de esos 
universos tuviesen las mismas leyes f\'\i sicas que el nuestro
(presumiblemente la mitad ser\'\i an de materia y la otra mitad de antimateria), 
y tambi\'en ser\'\i a de esperar que muchas de las civilizaciones avanzadas
correspondientes dominaran las t\'ecnicas de viajar o `saltar' a 
trav\'es de las dimensiones extra, al menos a trav\'es de algunas de 
ellas. En consecuencia se abrir\'\i an enormes posibilidades respecto a la 
expansi\'on de civilizaciones avanzadas simult\'aneamente a trav\'es de 
varios universos paralelos con las mismas leyes f\'\i sicas, dando 
lugar, como resultado, a imperios multidimensionales. Podr\'\i a
incluso suceder que la expansi\'on a otras galaxias paralelas a trav\'es
de dimensiones extra fuera m\'as f\'acil, con m\'as bajo costo, que la
expansi\'on dentro de la propia galaxia\footnote{Los primeros cient\'\i ficos 
que consideraron dimensiones extra y universos paralelos fueron,
probablemente, Maxwell y Faraday en el siglo XIX. Fuera de la esfera
cient\'\i fica esta idea tiene muchos miles de a\~nos de antig\"uedad.
En el presente estamos todav\'\i a en una fase muy prematura en el
estudio de los universos branas y no sabemos si estas ideas son en 
verdad realistas. Cumrun Vafa, de la Universidad de Harvard, opina que
el hecho de que no veamos alien\'\i genas a nuestro alrededor podr\'\i a
ser la primera prueba de la existencia de universos branas: todos los
alien\'\i genas avanzados habr\'\i an emigrado a universos paralelos
mejores que \'este (nuestro Universo tendr\'\i a 
`medida matem\'atica' cero) [3].}.    
 
En muchos otros universos, sin embargo, las leyes f\'\i sicas 
ser\'\i an diferentes, correspondientes quiz\'as a diferentes soluciones 
de la Teor\'\i a de Todo \'ultima (si es que esta Teor\'\i a existe,
independientemente de que nosotros podamos o no llegar a descubrirla). 
Estas soluciones dar\'\i an lugar, 
probablemente, a universos de `materia-sombra' respecto al nuestro,
la cual s\'olo interaccionar\'\i a con nuestra materia 
gravitacionalmente, en el caso en que fuese tra\'\i da a nuestro 
Universo mediante una tecnolog\'\i a apropiada. Esto no significa,
no obstante, que los universos-sombra estar\'\i an necesariamente faltos
de seres inteligentes. Si algunos de ellos tuvieran civilizaciones 
avanzadas, sus individuos podr\'\i an quiz\'as `saltar' a
nuestro Universo, pero no para colonizarlo, ya que ni siquiera ver\'\i an
nuestros planetas y estrellas, que atravesar\'\i an sin darse ni cuenta
(\'unicamente notar\'\i an la atracci\'on gravitatoria hacia sus centros). 
Y a la inversa, nosotros no podr\'\i amos ni ver, ni hablar
a los visitantes-sombra, aunque ellos quiz\'as podr\'\i an
intentar comunicarse con los posibles seres inteligentes de nuestro
Universo, a trav\'es de ondas gravitatorias por ejemplo. Respecto a los
universos de antimateria, los antiobservadores inteligentes tampoco
enviar\'\i an colonizadores\footnote{aunque podr\'\i an enviar antiprisioneros, 
cuya llegada se conocer\'\i a como explosiones de rayos gamma.}.

Tambi\'en tenemos que puntualizar que en [1] se hace uso continuo,
repetido del concepto de `observador inteligente' sin dar una definici\'on
de su significado, lo cual dificulta seguir los argumentos y c\'omputos
con propiedad. Por ejemplo, \que los hombres de Cro-Magnon y de
Neandertal contar\'\i an como civilizaciones de observadores inteligentes?
\que Y los seres humanos muy primitivos que habitan algunas selvas hoy en 
d\'\i a?  \que Pertenecen a la civilizaci\'on del planeta Tierra aunque 
sepan muy poco de ella?

En la discusi\'on que sigue en las pr\'oximas secciones usaremos las
siguientes definiciones intuitivas: 

{\it Civilizaciones primitivas}: Son aquellas civilizaciones que hacen 
un uso notable de tecnolog\'\i a en la vida diaria, pero no pueden
salir de sus planetas para colonizar otros en sistemas estelares
diferentes. Su conocimiento cient\'\i fico puede alcanzar muchos niveles
diferentes, yendo desde cero hasta niveles notablemente altos. En nuestro
planeta parece que s\'olo entrar\'\i an en el c\'omputo de civilizaciones 
primitivas los grupos de seres humanos de, aproximadamente, los \'ultimos 20.000
a\~nos, correspondientes a lo que los antrop\'ologos denominan el Hombre 
Moderno, no as\'\i\ los grupos pertenecientes a las versiones varias del
Hombre Primitivo, que s\'olo contar\'\i an como {\it civilizaciones muy 
primitivas}. Llamaremos a los individuos de las civilizaciones primitivas
{\it observadores inteligentes primitivos}.  

{\it Civilizaciones avanzadas}: Son aquellas civilizaciones tecnol\'ogicamente
capaces de co\-lonizar otros planetas en sistemas estelares diferentes del suyo, 
desde unos pocos planetas hasta miles de ellos o m\'as en el caso de 
{\it civilizaciones muy avanzadas}. Dependiendo de su nivel tecnol\'ogico, 
podr\'\i an incluso viajar a trav\'es de dimensiones extra (si es que 
existen), por lo que quiz\'as podr\'\i an visitar y colonizar planetas 
situados en algunas galaxias cercanas pertenecientes a posibles universos 
paralelos. Llamaremos a los individuos de estas civilizaciones {\it observadores 
inteligentes avanzados} y {\it muy avanzados}, respectivamente.

\section{Ideas Principales}\lvm  

Vamos a discutir en detalle la posibilidad de que nuestra peque\~na
civilizaci\'on terrestre est\'e inmersa en una civilizaci\'on grande 
sin saberlo. Esto nos conducir\'a de forma muy natural a la propuesta
de dos ideas principales que llamamos el `Principio Subantr\'opico'
y la `Conjetura de Indetectabilidad'.
 
Para comenzar volvamos al argumento principal. En nuestra galaxia
hay miles de millones de estrellas mucho m\'as antiguas que el Sol, siendo
muchas de ellas miles de millones de a\~nos m\'as antiguas, de hecho.
As\'\i\ pu\'es, parece de lo m\'as natural esperar, sin necesidad de
invocar la inflaci\'on cosmol\'ogica, que en una cantidad razonable
de sistemas estelares deben haber aparecido civilizaciones 
tecnol\'ogicas y una fracci\'on de ellas (aunque sea 
peque\~na) debe haber sobrevivido lo suficiente para extenderse por,
al menos, amplias regiones de la galaxia. Es por tanto muy notable
el hecho de que el Sistema Solar no haya sido nunca abordado
ni colonizado por ninguna civilizaci\'on avanzada,.... \que o lo ha sido?

En nuestra opini\'on, hay un error importante en las suposiciones
(impl\'\i citas) de Olum respecto a las relaciones entre las 
diferentes civilizaciones puestas en contacto en el proceso de
expansi\'on. Aunque no menciona este punto tan crucial, da la
impresi\'on de que \'el cree que las civilizaciones m\'as avanzadas
`tiran' de las menos avanzadas hasta su propio nivel para integrarlas,
o al contrario, las explotan, perjudican o aniquilan para conquistar
su planeta, en el caso de colonizadores agresivos. Estamos
totalmente de acuerdo en que las civilizaciones avanzadas agresivas
explotar\'\i an/perjudicar\'\i an/aniquilar\'\i an a las menos avanzadas
tanto como fuera conveniente para ellas. En el caso de \civ \ava
no agresivas, sin embargo, la posibilidad de que integraran a las menos 
avanzadas s\'olo tendr\'\i a sentido si \'estas no fuesen muy inferiores.
Es decir, si el salto o distancia entre las dos \civ no fuera muy grande,
entonces ser\'\i a realista esperar que la civilizaci\'on superior empujase
o tirase de la inferior hacia su propio nivel, al menos en cierta medida.
En algunos casos, sin embargo, las \civ \ava no agresivas encontrar\'\i an
planetas con \civ \pri o muy primitivas, con una enorme distancia 
(tecnol\'ogica, cient\'\i fica y gen\'etica) entre ellas. En particular,
las diferencias entre sus capacidades cerebrales y las de los
individuos primitivos podr\'\i an ser pat\'eticas. En estas circunstancias, 
ser\'\i a completamente irrealista e ingenuo esperar que los individuos 
avanzados intentaran integrar a los primitivos en sus propias 
civilizaciones. Lo que ser\'\i a de esperar, si acaso, es que se comportasen
de forma `ecol\'ogica' hacia ellos, trat\'andolos de manera similar
a una especie protegida, sin interferir (o s\'olo muy discretamente)
con su evoluci\'on natural.      
  
Con esta percepci\'on es ahora mucho m\'as f\'acil de aceptar
la posibilidad de que el Sistema Solar pudiera haber sido hallado
o colonizado hace muchos miles, o incluso millones de a\~nos, por
al menos una civilizaci\'on avanzada no agresiva que habr\'\i a tratado, 
y pudiera que todav\'\i a tratara, a nuestro planeta como una reserva natural
protegida. De hecho, incluso podr\'\i an haber tra\'\i do muchas
plantas y animales a la Tierra, incluyendo a nuestros ancestros,
presumiblemente para mejorar sus condiciones de vida (pudieran
haber estado en peligro de extinci\'on en su planeta de origen,
por ejemplo)\footnote{Una de las actividades esperables de las
\civ \ava ser\'\i a la diseminaci\'on de la vida en planetas 
`prometedores', del mismo modo que nosotros plantamos \'arboles
en lugares adecuados. Si el comienzo de la vida sobre la Tierra hubiera
sucedido de esta manera, entonces todos los seres vivos terrestres
tendr\'\i amos bloques o bases de ADN comunes con los seres vivos de
miles de otros planetas que hubiesen pasado por el mismo proceso de 
inseminaci\'on con las mismas bacterias. Es por tanto concebible que, bajo 
estas circunstancias, se pudieran haber tra\'\i do plantas y animales a la 
Tierra cuyo origen extraterrestre fuera imposible de detectar por ning\'un 
bi\'ologo o genetista.}. Quiz\'as el Sistema Solar haya sido visitado 
por colonizadores agresivos, tanto como por no agresivos, habi\'endose
producido algunas batallas como resultado, o habi\'endose llegado simplemente 
a negociaciones pac\'\i ficas entre ellos. Quiz\'as los perdedores 
agresivos volver\'an en el futuro, para intentarlo otra vez....

Esta visi\'on sobre nosotros mismos, una peque\~na 
civilizaci\'on primitiva inmersa en una gran civilizaci\'on avanzada,
nos lleva directamente a hacernos conscientes de que pudi\'eramos
no encontrarnos entre los observadores inteligentes t\'\i picos
de nuestra galaxia, sino s\'olo entre una peque\~na  proporci\'on 
de observadores inteligentes primitivos, en su lugar, ignorantes 
por completo de su bajo nivel. Los observadores inteligentes
t\'\i picos ser\'\i an los ciudadanos de las \civ \ava y muy avanzadas,
quienes ser\'\i an los `due\~nos' de la galaxia. Pero nuestra galaxia 
es s\'olo una galaxia t\'\i pica de nuestro Universo observable. Esto
nos conduce de forma muy natural a nuestra primera propuesta:

{\it El Principio Subantr\'opico}: Nosotros no somos t\'\i picos 
entre los observadores inteligentes del Universo. Las \civ t\'\i picas
de las galaxias t\'\i picas est\'an cientos de miles, o millones, de
a\~nos m\'as evolucionadas que la nuestra y, consecuentemente, los
observadores inteligentes t\'\i picos son \'ordenes de magnitud
m\'as inteligentes que nosotros.

Obs\'ervese que el Principio Subantr\'opico es casi equivalente a
la propuesta de que, en el presente, todas las galaxias t\'\i picas
del Universo (o vastas regiones de las mismas) est\'an ya colonizadas
por \civ avanzadas, o muy avanzadas, en las que una peque\~na proporci\'on
de sus individuos pertenecen a subcivilizaciones primitivas, como la nuestra.
El que las subcivilizaciones \pri conozcan o ignoren su
bajo nivel depender\'\i a, muy probablemente, de los est\'andares
\'eticos de las \civ \ava en la que estuvieran inmersas.
Si los est\'andares fuesen bajos, los individuos de las subcivilizaciones
primitivas ser\'\i an maltratados de muchas formas, con toda certeza, de la
misma manera que en nuestra civilizaci\'on grupos numerosos de seres
humanos maltratan a otros seres humanos m\'as d\'ebiles y/o en 
inferioridad de condiciones, as\'\i\ como a los animales en ge\-neral.
En este caso, pu\'es, los individuos primitivos ser\'\i an penosamente
conscientes de su bajo nivel. Si, por el contrario, los est\'andares
\'eticos de los individuos avanzados fueran elevados, entonces muy
probablemente respetar\'\i an la evoluci\'on natural (biol\'ogica, social
y cultural) de las subcivilizaciones primitivas trat\'andolas 
`ecol\'ogicamente' de forma si\-milar a cualquier especie protegida.
As\'\i\ que en este caso, que creemos podr\'\i a muy bien describir la 
situaci\'on de la civilizaci\'on terrestre, los individuos primitivos 
ser\'\i an completamente inconscientes de la existencia
de la civilizaci\'on avanzada grande en la que estar\'\i an inmersos.

Ahora viene una observaci\'on importante: si el Sistema Solar formara
parte del territorio de una civilizaci\'on avanzada, entonces \que por qu\'e 
no se detectan se\~nales de civilizaci\'on en ninguno de los 
planetas s\'olidos y sat\'elites grandes que se hallan en su interior?
Ser\'\i a de lo m\'as natural que se hubiesen constru\'\i do bases a 
todo lo largo y ancho del Sistema Solar (incluyendo bases subterr\'aneas
y submarinas en el planeta Tierra) y quiz\'as algunas colonias
sobre, o debajo de, la superficie de algunos planetas s\'olidos
y de algunos sat\'elites grandes (\exc esto es exactamente lo que nosotros
planeamos hacer en el futuro!). La respuesta m\'as simple ser\'\i a que
`ellos' no encuentran el Sistema Solar lo suficientemente atractivo como
para establecerse en \'el y, en consecuencia, s\'olo tienen unas pocas bases
diminutas dif\'\i ciles de detectar. Sin embargo, independientemente
de si encuentran el Sistema Solar atractivo o no para establecerse y levantar
colonias, creemos que todas las \civ \ava tienen que ser conscientes,
necesariamente, de la existencia de \civ \ava {\it agresivas} y, como
resultado, deben haber desarrollado sistemas de camuflaje muy
sofisticados, de manera que ning\'un observador externo (ni sus sondas)
pueda detectar se\~nal alguna de civilizaci\'on. Es probable que, en 
muchos casos, incluso manipulen y distorsionen los datos globales de
sus planetas (temperatura, composici\'on del aire, etc.), para 
confundir a los observadores externos, con el prop\'osito de 
disuasi\'on\footnote{Puede sonar extra\~no que las \civ \ava
necesiten protegerse contra agresores. Sin embargo, no hay una 
s\'ola prueba o indicaci\'on de que el desarrollo \'etico de una
civilizaci\'on, o un individuo, crezca en paralelo con su nivel de 
bienestar material o con su desarrollo tecnol\'ogico y cient\'\i fico.
Tambi\'en puede argumentarse que las \civ \ava agresivas tienen que
autoaniquilarse, lo que parece una suposici\'on razonable. La cuesti\'on
crucial no es, sin embargo, si se autoaniquilar\'an o no, sino {\it cu\'anto 
da\~no} pueden hacer a otras \civ (tanto \pri como avanzadas) antes de 
autoaniquilarse.}. Este es el contenido de nuestra segunda propuesta:
 
{\it La Conjetura de Indetectabilidad}: Gen\'ericamente, todas las
\civ suficientemente \ava camuflan sus planetas por razones de seguridad,
de manera que ninguna se\~nal de civilizaci\'on pueda ser detectada
por observadores externos, quienes s\'olo obtendr\'\i an datos
distorsionados con el prop\'osito de disuasi\'on.  
 
Obs\'ervese que si esta conjetura resultara ser cierta, entonces no
podr\'\i amos estar seguros ni siquiera de que la civilizaci\'on terrestre
sea la \'unica civilizaci\'on que habita en el Sistema Solar, como creemos
firmemente (esto es independiente, de hecho, de si nuestra civilizaci\'on
est\'a o no inmersa en una civilizacion avanzada grande, s\'olo es
necesario que tengamos `vecinos avanzados'). De hecho, resulta notable
la inconsistencia en el razona\-miento cient\'\i fico que se usa en
las observaciones astron\'omicas de planetas y sat\'elites.  Se utiliza
como punto de partida la suposici\'on, no probada, de que en
el origen de las observaciones no hay seres inteligentes manipulando 
los datos que recibimos, y luego se concluye que no hay se\~nales de 
vida inteligente {\it como prueban los datos}. Pero esta suposici\'on 
podr\'\i a no ser cierta. Lo correcto en este caso ser\'\i a concluir 
que no hay se\~nales de {\it civilizaciones primitivas}, como la
nuestra, que se dejar\'\i an detectar por observadores externos, pero
{\it nada} puede decirse respecto a la posibilidad de {\it \civ 
avanzadas}, capaces de confundir y despistar a nuestros telescopios, 
detectores y sondas, y que no se dejar\'\i an detectar.    

Finalmente, tenemos que mencionar que el primer erudito,
al menos en la historia occidental, que sugiri\'o que muchas estrellas
ah\'\i\ fuera podr\'\i an tener planetas similares al nuestro: con
plantas, animales, gente, etc., fue Giordano Bruno, en el siglo XVI.
Afirmaba que el Sol s\'olo era una estrella entre mir\'\i adas,
y por lo tanto, como el Sol, muchas otras estrellas tendr\'\i an 
tambi\'en planetas girando a su alrededor y seres vivos los 
habitar\'\i an [4]. Para apreciar el genio de Giordano Bruno,
hay que tener en cuenta que vivi\'o en una \'epoca en la que m\'as 
del 99\% de los intelectuales cre\'\i an que la Tierra era el centro
del Universo, y unos pocos, como Cop\'ernico y Galileo, cre\'\i an
que era el Sol el centro del Universo, en su lugar, siendo las
estrellas cuerpos celestiales brillantes de naturaleza 
desconocida\footnote{Por \'estas y otras ideas Giordano Bruno fue
hecho prisionero durante ocho a\~nos y finalmente fue quemado en 
la hoguera en Roma, en la piazza Campo di Fiori, el 17 de Febrero de
1600. La iglesia cat\'olica, que hace algunos a\~nos se disculp\'o
por el trato dado a Galileo, nunca se ha disculpado, sin embargo, 
por el trato que infligi\'o a Giordano Bruno.}. Hoy en d\'\i a sabemos 
que el Universo no tiene centro y que nuestro planeta es s\'olo una
part\'\i cula de polvo diminuta en su inmensidad. A pesar de esto,
para muchos seres humanos la Tierra es a\'un el centro del Universo,
el planeta `elegido' habitado por los seres m\'as perfectos e
inteligentes de todo el Universo: la Corona de la Creaci\'on.
(\exc Existen incluso cient\'\i ficos e intelectuales `normales'
que se preguntan si todo el Universo habr\'a sido creado s\'olo 
para que existamos nosotros, los seres humanos terrestres!).

\vskip 0.5cm  
  
\section{Conclusiones y Comentarios Finales}  

Hemos discutido la posibilidad de que nuestra civilizaci\'on pudiera
estar inmersa en una civilizaci\'on avanzada grande extendi\'endose
por (al menos) una regi\'on amplia de nuestra galaxia. Esto ser\'\i a
de esperar, de hecho, ya que en nuestra galaxia existen muchos miles
de millones de estrellas mucho m\'as antiguas que el Sol. Haciendo dos 
suposiciones simples y naturales vemos que esta posibilidad no puede
descartarse. 

La primera suposici\'on explicar\'\i a por qu\'e los miembros o ciudadanos 
de la civilizaci\'on grande no interaccionar\'\i an ni socializar\'\i an
con nosotros (abierta y oficialmente, al menos). La raz\'on ser\'\i a que 
nosotros no cumplimos los requisitos m\'\i nimos para ser miembros  ni 
asociados, aunque pudiera suceder que nos considerasen como posibles mascotas
o `amigos'. Generalizando esta situaci\'on, teniendo en cuenta que vivimos en
una galaxia t\'\i pica, llegamos al Principio Subantr\'opico, que dice que 
nosotros no somos t\'\i picos entre los observadores inteligentes del 
Universo, sino que nos encontramos muy por debajo de los est\'andares.    

La segunda suposici\'on, que llamamos la Conjetura de la Indetectabilidad,
explicar\'\i a por qu\'e no detectamos se\~nal alguna de esta 
civilizaci\'on grande en la que estar\'\i amos inmersos. La raz\'on
ser\'\i a que, en general, todas las civilizaciones avanzadas ser\'\i an
indetectables por razones de seguridad, debido a la existencia de
\civ \ava {\it agresivas}. En cualquier caso, \que por qu\'e raz\'on una
civilizaci\'on avanzada permitir\'\i a a cualquier civilizaci\'on
alien\'\i gena ver sus ciudades, laboratorios, instalaciones militares, 
etc. cuando, en su lugar, podr\'\i a confundirla y despistarla muy f\'acilmente?

El Principio Subantr\'opico es casi equivalente a la propuesta de que
todas las galaxias t\'\i picas del Universo est\'an ya colonizadas
(o al menos amplias regiones de las mismas) por \civ \ava o muy avanzadas, 
lo cual es una suposici\'on de lo m\'as natural teniendo en cuenta que 
muchos miles de millones de estrellas que pueblan las galaxias t\'\i picas 
son miles de millones de a\~nos m\'as antiguas que el Sol. En estas \civ \ava 
grandes siempre existir\'\i an, de forma gen\'erica, un peque\~no porcentaje
de individuos que pertenecer\'\i an a subcivilizaciones primitivas. Si
los est\'andares \'eticos de los individuos avanzados fuesen bajos, entonces 
los individuos primitivos ser\'\i an maltratados de muchas maneras (quiz\'as
incluso aniquilados). Si los est\'andares \'eticos de los individuos 
avanzados fueran elevados, en su lugar, entonces probablemente
tratar\'\i an a los individuos primitivos de una manera ecol\'ogica;
es decir, como a una especie protegida que vive en una reserva natural.
En este caso, que bien pudiera describir la situaci\'on de nuestra 
civilizaci\'on, la mayor\'\i a de los individuos primitivos ignorar\'\i an
completamente la existencia de la civilizaci\'on avanzada en la que
estar\'\i an inmersos.

Tambi\'en hemos argumentado que la idea de universos branas, aunque
a\'un se encuentra en una fase muy prematura, podr\'\i a de hecho agravar 
enormemente el problema del `alien\'\i gena ausente', puesto de manifiesto
por primera vez por Enrico Fermi, como ya hemos mencionado.  
La raz\'on es que, si existieran otros 
universos paralelos con las mismas leyes f\'\i sicas que el nuestro,
podr\'\i a ocurrir que algunas de sus civilizaciones avanzadas 
fueran t\'ecnicamente capaces de `saltar' a trav\'es de las dimensiones
extra a nuestra galaxia con el prop\'osito de expansi\'on y colonizaci\'on.
Como resultado, podr\'\i a incluso suceder que los due\~nos del Sistema 
Solar (si es que existen) hubieran venido de otro universo y hubiesen creado
un imperio gigantesco multidimensional, con amplias regiones de territorio
en varias galaxias `paralelas'. Podr\'\i a ocurrir tambi\'en que las
\civ \ava encontraran m\'as eficiente (m\'as barato y preferible
energ\'eticamente) expandirse a trav\'es de dimensiones extra que
dentro de su propia galaxia.

Finalmente, en el Ap\'endice tratamos el tema de los posibles contactos
e interacciones entre \civ o individuos avanzados y \civ o individuos 
primitivos. En nuestra opini\'on, ser\'\i a altamente improbable que 
una civilizaci\'on avanzada no agresiva se presentase abiertamente 
a alguna civilizaci\'on primitiva. No obstante, a nivel de los
individuos hemos identificado tres causas o razones principales 
que podr\'\i an motivar a individuos de \civ \ava a interaccionar
o buscar relaciones con individuos primitivos: 
objetivos/fines cient\'\i ficos, afecto/entretenimiento y prop\'ositos 
delictivos de todo tipo. Tambi\'en hacemos notar que el Principio 
Subantr\'opico y la Conjetura de Indetectabilidad predicen una 
probabilidad muy baja de \'exito para el proyecto SETI, debido al peque\~no 
porcentaje de civilizaciones tecnol\'ogicas que ser\'\i an susceptibles
de ser detectadas (el per\'\i odo de detectabilidad de una civilizaci\'on
promedio podr\'\i a durar menos de 500 a\~nos).

\vskip .8cm

\centerline{\large \bf Ap\'endice}

\vskip .6cm

En lo que sigue discutiremos las posibles fuentes de contacto e
interacciones entre \civ o individuos avanzados y \civ o individuos
primitivos. Como argumentamos en los preliminares, creemos muy improbable
el que una civilizaci\'on avanzada no agresiva contactase a alguna 
civilizaci\'on primitiva `abierta y oficialmente', 
al menos hasta que esta \'ultima alcanzase un grado 
de desarrollo notable que nuestra civilizaci\'on no
ha alcanzado a\'un. Las \civ \ava agresivas, sin embargo, se
`presentar\'\i an' antes, despu\'es o durante el ataque, seg\'un su
propia conveniencia. (El hecho de que nuestra civilizaci\'on no haya 
sido nunca atacada por alien\'\i genas agresivos, hasta lo que la
historia conoce, podr\'\i a ser en efecto un indicio de que
pertenecemos a una civilizaci\'on avanzada no agresiva que protege 
el planeta Tierra como parte de su territorio.).

Si ahora consideramos posibles contactos y relaciones entre individuos
de \civ \ava e individuos primitivos, en vez de entre sus civilizaciones,
se presentan muchas m\'as posibilidades. Con el objeto de
identificar qu\'e individuos avanzados podr\'\i an
estar interesados en interaccionar y relacionarse con individuos
primitivos, y por qu\'e razones, llegamos a distinguir tres fuentes
principales de contactos:  

1) Investigaci\'on cient\'\i fica llevada a cabo por cient\'\i ficos 
oficiales relacionados con las ciencias de la vida, tales como bi\'ologos, 
m\'edicos, antrop\'ologos, soci\'ologos, sic\'ologos, etc. El que las 
correspondientes actividades de investigaci\'on pudieran da\~nar 
a los individuos primitivos (f\'\i sica o mentalmente), depender\'\i a 
de las regulaciones legales de las \civ \ava con respecto al trato 
\'etico hacia individuos de \civ primitivas.

2) Entretenimiento, afecto, etc. Es decir, un individuo de una
civilizaci\'on avanzada podr\'\i a establecer contacto con individuos
primitivos simplemente para entretenerse y relajarse. El individuo
avanzado podr\'\i a tener, con respecto a los individuos primitivos,
el tipo de sentimientos que nos empujan a nosotros a interaccionar
y jugar con perros y gatos y otras muchas especies. Adem\'as,
si en nuestro planeta hay millones de amantes de los perros y
millones de amantes de los gatos, y hay incluso amantes de las serpientes,
de los cerdos, .... y de los gorilas, ser\'\i a de lo m\'as natural esperar 
que pudieran existir algunos amantes de los individuos primitivos, 
en particular de los humanos terrestres, entre los 
alien\'\i genas avanzados. \que Por qu\'e no? Podr\'\i a darse el caso,
especialmente entre aquellos individuos avanzados que tuvieran que pasar 
largas temporadas trabajando en planetas primitivos, viviendo en 
bases aburridas subterr\'aneas o submarinas, que existir\'\i an en
nuestro planeta con toda seguridad en el caso en que nuestra 
civilizaci\'on estuviera inmersa en una civilizaci\'on grande (los
trabajadores de las bases ser\'\i an los `guardas' o militares que
estar\'\i an a cargo del planeta).

3) Prop\'ositos delictivos de todo tipo, incluyendo actividades
llevadas a cabo por cient\'\i ficos oficiales que estuviesen 
prohibidas por sus regulaciones \'eticas legales. Podemos
imaginarnos docenas de prop\'ositos delictivos diferentes por los
que los individuos primitivos podr\'\i an ser secuestrados, torturados
e incluso asesinados, incluyendo t\'opicos abyectos tales como
`alta gastronom\'\i a' y juegos s\'adicos. Siendo realistas, s\'olo
hay que pensar en el trato cruel que algunos seres humanos infligen
a sus v\'\i ctimas, ya sean otros seres humanos (a menudo ni\~nos) o
animales. La cuesti\'on es que el nivel \'etico de un individuo,
o una civilizaci\'on, no crece necesariamente en paralelo con sus
logros cient\'\i ficos y tecnol\'ogicos, ni con su nivel de bienestar
material. En el caso en que nuestra civilizaci\'on estuviera inmersa
en una civilizaci\'on grande, una de las tareas de los `guardas'  
viviendo en las bases ser\'\i a, sin duda, ahuyentar
a los cazadores de humanos y a otros proscritos.  
  
\vskip 0.2cm 

En cuanto al proyecto SETI: search for extraterrestrial intelligence 
(b\'usqueda de inteligencia extraterrestre), si la Conjetura de 
Indetectabilidad resultara ser cierta, entonces SETI pasar\'\i a a
ser SETPI: search for extraterrestrial {\it primitive} intelligence
(b\'usqueda de inteligencia extraterrestre {\it primitiva}).
La raz\'on ser\'\i a que, en este caso, s\'olo las civilizaciones
primitivas podr\'\i an ser detectadas por observadores externos. 
Por otro lado, si el Principio Subantr\'opico es correcto, entonces
las \civ \pri ser\'\i an muy escasas comparadas con el total de las
civilizaciones tecnol\'ogicas, y mucho m\'as escasas ser\'\i an
aquellas con un nivel tecnol\'ogico apropiado para producir emisiones 
electromagn\'eticas que pudieran ser detectadas por \civ distantes. 
(En nuestra opini\'on, el per\'\i odo de `detectabilidad' de una
civilizaci\'on promedio podr\'\i a durar menos de 500 a\~nos). As\'\i\
pu\'es, la probabilidad de que una civilizaci\'on primitiva 
detectara otra ser\'\i a muy peque\~na. Por estas razones, el
Principio Subantr\'opico y la Conjetura de Indetectabilidad predicen
una baja probabilidad de \'exito para el proyecto SETI\footnote{Los expertos
del proyecto SETI deber\'\i an quiz\'as unirse a la competencia para
colaborar: las `antenas humanas' o `expertos-SETI-alternativos', que
afirman mantener contactos estables y a largo plazo con `partners'
alien\'\i genas, algunos desde su casa; es decir, desde sus propios 
planetas y otros desde bases subterr\'aneas o submarinas aqu\'\i\ en
la Tierra, donde vivir\'\i an temporalmente. El record en tales supuestas
relaciones lo ostenta, probablemente, el grupo espa\~nol Aztl\'an,
quienes llevan alrededor de {\it veinticinco a\~nos} reuni\'endose una
noche por semana para establecer (lo que ellos afirman ser) comunicaciones
telep\'aticas con un grupo de soci\'ologos del planeta Apu, en \'orbita
alrededor de Alfa B Centauro, que se encontrar\'\i an en su planeta.}.

Un \'ultimo comentario es que nunca hemos hecho ning\'un tipo de 
investigaci\'on en el tema de los contactos declarados como 
alien\'\i genas. As\'\i\ pues, no tenemos ninguna opini\'on sobre
la veracidad o falsedad de los supuestos contactos que circulan 
por los medios de comunicaci\'on. No obstante, creemos que debe
de ser imposible identificar contactos alien\'\i genas verdaderos 
(si es que existen) simplemente leyendo los informes dados a, o
escritos por, sus contactados. La raz\'on es que,
para nuestra intuici\'on, las afirmaciones de \civ  mucho m\'as
avanzadas que nosotros tienen que sonar {\it necesariamente}
rid\'\i culas, hilarantes, locuras de pura ciencia ficci\'on. Pero
lo mismo habr\'\i a sucedido si nosotros hubi\'eramos descrito 
nuestros aparatos de televisi\'on, nuestros aviones, nuestros hornos
microondas, nuestros ordenadores, etc .... a la gente de \exc hace 
s\'olo 100 a\~nos! Queremos hacer notar tambi\'en que muchas
personas, incluyendo muchos cient\'\i ficos, tienen gran
reluctancia y aversi\'on, muy profundamente enraizadas, en aceptar 
la posibilidad de que puedan existir especies extraterrestres
mucho m\'as avanzadas e inteligentes que nosotros, que incluso
pudieran visitar nuestro planeta. Llamamos a este prejuicio el
`S\'\i ndrome de la Corona de la Creaci\'on' (SCC), por razones
obvias. Curiosamente, mientras que muchas personas religiosas no
sufren el SCC, muchos ateos lo sufren (una explicaci\'on podr\'\i a
ser que crecieron en familias muy religiosas que implantaron en las
mentes de sus hijos impresiones muy fuertes sobre la grandeza y
lo excepcional de la especie humana)\footnote{En resumen, los afectados
por el SCC parecen creer firmemente, o esperar, que nadie en 
todo el Universo puede hacer lo que nosotros no podemos hacer, en
particular viajes interestelares. Es interesante observar tambi\'en que 
entre ellos abundan los que albergan grandes expectativas 
sobre las capacidades y los logros futuros que nos esperan a nuestra 
civilizaci\'on. Sin embargo, en sus razonamientos parecen carecer de 
la habilidad de intercambiar `nosotros' por `ellos' y el 
futuro por el pasado, en referencia a posibles \civ millones de a\~nos 
m\'as antiguas que la nuestra. Por ejemplo, muchos de ellos aceptan
de buen grado las sugerencias del tipo `llegaremos a otros planetas y 
estrellas', `colonizaremos la galaxia', etc. y sin embargo no 
pueden ni escuchar las sugerencias de que `ellos pueden haber llegado a otros
planetas y estrellas (inclu\'\i dos el Sistema Solar y la Tierra)',
`ellos pueden haber colonizado la galaxia', etc., que son recibidas con
sarcasmo, desaprobaci\'on e incluso agresividad.}.

\vskip .15in

Para terminar, queremos se\~nalar que la situaci\'on actual en cuanto 
a la b\'usqueda de inteligencia extraterrestre (Search for ExtraTerrestrial
Intelligence) pudiera muy bi\'en venir descrita por la popular canci\'on 
protesta americana de los a\~nos sesenta, que transcribimos a continuaci\'on:

\vskip .20in

{\it {\ \ \ } Where have all the aliens gone?} 

{\it {\ \ \ } Long time passing....}

{\it {\ \ \ } Where have all the aliens gone?}

{\it {\ \ \ } Long time ago....}

{\it {\ \ \ } Where have all the aliens gone?}

{\it {\ \ \ } Could be hidden everywhere!}

{\it {\ \ \ } When will we ever learn?}

{\it {\ \ \ } When will we ever learn?}

{\ \ \ \ }(Repetir tres veces)

\vskip .4in

\noi
\centerline{\large \bf Agradecimientos}

\vskip .7cm

Doy las gracias a los muchos lectores de este art\'\i culo que me han 
expresado su apreciaci\'on por el mismo. Desde que se public\'o en los
archivos cient\'\i ficos `Popular Physics' y `Space Physics' en Agosto
de 2003, hace exactamente un a\~no, he recibido una avalancha de cuestiones,
comentarios, sugerencias, observaciones,... y tambi\'en una cantidad 
apreciable de informaci\'on en la forma de bibliograf\'\i a y de p\'aginas 
web, relacionadas con las materias que se discuten en este art\'\i culo.
Desafortunadamente no he tenido oportunidad de leer m\'as que una m\'\i nima 
parte de esa informci\'on, por falta de tiempo. Tambi\'en estoy muy 
agradecida a mi amiga Mar\'\i a Teresa Fern\'andez Mart\'\i nez por su ayuda 
inestimable en la traducci\'on de este art\'\i culo al castellano y por 
muchas conversaciones muy interesantes sobre temas relacionados con el mismo.

\vskip .4in 

\noi
{\large \bf Referencias}
  
\vskip .2in  
   
\noi  
[1] K. Olum, `Conflict between anthropic reasoning and  
observation', gr-qc/0303070.  

\noi 
[2] N. Arkani-Hamed, S. Dimopoulos and G.R. Dvali, 
Phys. Lett. B429, 263, 1998; Phys. Rev. D 59, 86004, 1999;
Phys. Today 55N2, 35, 2002.\\
L. Randall and R. Sundrum, Phys. Rev. Lett. 83, 4690, 1999.

\noi 
[3] C. Vafa, private communication.

\noi
[4] G. Bruno, {\it On the Infinite Universe and Worlds}, 1584.

\vskip .4in

\noi
{\large \bf Nota Biogr\'afica}

\vskip .6cm

La autora, Beatriz Gato Rivera, es natural de Madrid. Licenciada y
doctora en Ciencias F\'\i sicas por la Universidad Complutense de
Madrid, es especialista en F\'\i sica de Part\'\i culas Elementales
y F\'\i sica Matem\'atica. Habiendo pasado tres a\~nos postdoctorales
en el MIT (Instituto de Tecnolog\'\i a de Masachussets) y otros tres 
a\~nos en el CERN, laboratorio europeo de investigaci\'on nuclear,
desde 1990 forma parte de la plantilla cient\'\i fica del CSIC
(Consejo Superior de Investigaciones Cient\'\i ficas). e-mail: t38@nikhef.nl

\vskip .3in